\documentclass[aps,prl,twocolumn,showpacs,amsmath,amssymb,floatfix,groupedaddress]{revtex4-1}

\usepackage{graphicx}
\usepackage{bm}
\usepackage{color}

\begin{document}

\title{Experimental Violation of Two-Party Leggett-Garg Inequalities \\with Semi-weak Measurements}

\author{J. Dressel}
\author{C. J. Broadbent}
\author{J. C. Howell}
\author{A. N. Jordan}
\affiliation{Department of Physics and Astronomy, University of Rochester, Rochester, New York 14627, USA}

\date{\today}

\newcommand{\mean}[1]{\langle #1 \rangle}           
\newcommand{\cmean}[2]{\,_{#1}\langle #2 \rangle}   

\newcommand{\ket}[1]{|#1\rangle}                    
\newcommand{\bra}[1]{\langle #1|}                   
\newcommand{\ipr}[2]{\langle #1 | #2 \rangle}       
\newcommand{\opr}[2]{\ket{#1}\bra{#2}}              
\newcommand{\pprj}[1]{\opr{#1}{#1}}                 

\newcommand{\Tr}[1]{\mbox{Tr}\left[#1\right]}       
\newcommand{\comm}[2]{\left[#1,\,#2\right]}         
\newcommand{\acomm}[2]{\left\{#1,\,#2\right\}}      
\def\R{\mbox{Re}}                                   
\newcommand{\op}[1]{\hat{#1}}                       
\def\prj{\op{\Pi}}                                  

\newcommand{\oper}[1]{\mathcal{#1}}                 
\newcommand{\prop}[1]{\textit{#1}}                  
\def\gbar{\bar{\gamma}}
\def\ebar{\bar{\eta}}

\begin{abstract}
  We generalize the derivation of Leggett-Garg inequalities to systematically treat a larger class of experimental situations by allowing multi-particle correlations, invasive detection, and ambiguous detector results.  Furthermore, we show how many such inequalities may be tested simultaneously with a single setup.  As a proof of principle, we violate several such two-particle inequalities with data obtained from a polarization-entangled biphoton state and a semi-weak polarization measurement based on Fresnel reflection.  We also point out a non-trivial connection between specific two-party Leggett-Garg inequality violations and convex sums of strange weak values.
\end{abstract}


\maketitle

To better understand and identify the apparent division between macroscopic and microscopic behavior, Leggett and Garg have distilled common implicit assumptions about the macroscopic world into a set of explicit postulates that they dub \emph{macrorealism} (MR)~\cite{Leggett}. From these postulates, they construct inequalities analogous to Bell inequalities~\cite{Bell} but involving multiple correlations in time.  Such Leggett-Garg inequalities (LGIs) must be satisfied by any theory compatible with MR, but may be violated by quantum mechanics.  As such, LGI violations have received increasing interest as signatures of distinctly quantum behavior in qubit implementations \cite{Ruskov2006,Jordan,Wilde2010}, and have been recently confirmed experimentally in both solid-state~\cite{Palacios2010} and optical systems~\cite{Goggin2010}. 

In this Letter, we demonstrate a technique for systematically deriving generalized LGIs that admit multiple parties, invasive detection, and/or ambiguous detector results by considering a specific two-particle experimental setup with three measurements.  We proceed to experimentally violate several such two-party LGIs simultaneously with a single data set produced from a setup using a \emph{semi-weak} polarization measurement on an entangled biphoton state.  The contextual values (CV) analysis of quantum measurement~\cite{Dressel2010} suggests a direct comparison between the classical and quantum treatments.  Finally, we show that specific two-party LGIs are equivalent to constraints on convex sums of conditioned averages (CA), which are the generalizations of the quantum weak value to an arbitrary measurement setup~\cite{Dressel2010,Aharonov1988}. The technique may be easily extended to check data from a setup with any number of measurements and parties.  

\emph{Generalized LGIs}.---A \emph{MR theory} consists of three key postulates:  (i) if an object has several distinguishable states available to it, then at any given time it is in only one of those states;  (ii) one can \emph{in principle} determine which state it is in without disturbing that state or its subsequent dynamics; and, (iii) its future state is determined causally by prior events~\cite{Leggett}.  Furthermore, we acknowledge that physical detectors may be imperfect by being (a) \emph{invasive} by altering the object state during the interaction, or (b) \emph{ambiguous} by reporting results that only correlate probabilistically with the object state due to inherent detector inefficiencies or errors.  

\begin{figure}[t]
\includegraphics[scale=.8]{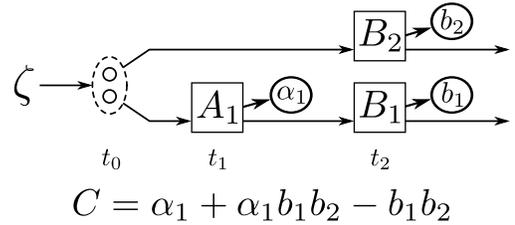}
\caption{MR measurement schematic.  An object pair is picked from an ensemble $\zeta$ at time $t_0$.  At $t_1$ object 1 of the pair interacts with an imperfect detector for the property $\prop{A}_1$, which reports a generalized value $\alpha_1$.  At $t_2$ both objects interact with unambiguous detectors for the properties $\prop{B}_1$ and $\prop{B}_2$ that report values $b_1$ and $b_2$.  The two-party LG correlation $C$ is constructed from the measured results.} \label{fig:leggett-garg}
\end{figure}

For convenience we consider dichotomic properties in what follows, though the discussion can be easily extended.  Unambiguous detector outcomes will be assigned the (arbitrary) values $\{-1,1\}$ corresponding to the two possible states of the property being measured.  Ambiguous detectors will be calibrated to report the same ensemble average as an unambiguous detector for the same property.  To do so, their outcomes must be assigned \emph{generalized values} $\alpha\in S$ from an expanded set $S$, with $\min S \leq -1$ and $\max S \geq 1$, to compensate for the imperfect state correlation of the outcomes.  Such generalized values are the classical equivalent of quantum CV~\cite{Dressel2010} and may be determined by measuring pure ensembles of either $\pm 1$. 

We now derive a specific two-party generalized LGI for a particular experimental setup, keeping in mind that the method may be extended to any setup.  Consider a pair of MR objects that interacts with a sequence of detectors as shown in Fig.~\ref{fig:leggett-garg}.  At time $t_0$ the pair is picked from a known ensemble $\zeta$.  At time $t_1$ object 1 of the pair interacts with an imperfect detector for the dichotomic property $\prop{A}_1$, which reports a generalized value $\alpha_1\in S_1$.  Finally, at time $t_2$ objects 1 and 2 interact with unambiguous detectors for the dichotomic properties $\prop{B}_1$ and $\prop{B}_2$, respectively, which report the values $b_1, b_2\in\{-1,1\}$.  

For each object pair, we can keep all three results to construct the correlation product $\alpha_1 b_1 b_2$, or we can ignore some results as non-selective measurements~\cite{Kraus1983} to construct the alternate quantities $\alpha_1$, $b_1$, $b_2$, $\alpha_1 b_1$, $\alpha_1 b_2$, or $b_1 b_2$.  Since the latter terms involve voluntary loss of information after the measurement has been performed, we can compute them all from the \emph{same data set}.  Exploiting this freedom, we construct the correlation $C = \alpha_1 + \alpha_1 b_1 b_2 - b_1 b_2$ for each measured pair, which lies in the range, $-|1-2\min S_1|\leq C\leq|2\max S_1-1|$.  

We repeat this procedure many times and average the results of $C$ to obtain, $\mean{C} = \sum_{\alpha_1,b_1,b_2} P(\alpha_1 | \zeta) P(b_1,b_2 | \zeta, \alpha_1) \left( \alpha_1 + \alpha_1 b_1 b_2 - b_1 b_2 \right)$, where $P(\alpha_1 | \zeta)$ is the probability of detecting $\alpha_1$ given the initial ensemble $\zeta$, and $P(b_1, b_2 | \zeta, \alpha_1)$ is the probability of detecting $b_1$ and $b_2$ given the initial ensemble $\zeta$ and the possibly invasive detection of $\alpha_1$.  

Generally, we cannot separate the sums due to the $\alpha_1$-dependence of $P(b_1,b_2 | \zeta, \alpha_1)$, so the best guaranteed bounds are $-|1-2\min S_1|\leq \mean{C}\leq|2\max S_1-1|$.  As a special case, if the detector for $\prop{A}_1$ is \emph{unambiguous} then $\min S_1 = -1$, $\max S_1 = 1$, and we find the LGI,
\begin{equation}
  -3 \leq \mean{\prop{A}_1 + \prop{A}_1\prop{B}_1\prop{B}_2 - \prop{B}_1\prop{B}_2} \leq 1.
  \label{eq:leggett-garg-inequality1}
\end{equation}

Alternatively, if the detector is \emph{noninvasive} so that $P(b_1, b_2 | \zeta, \alpha_1) = P(b_1, b_2 | \zeta)$ then the sums do separate and we can average $\prop{A}_1$ first to find, $\mean{C} = \sum_{b_1,b_2}P(b_1,b_2 | \zeta)\left( \mean{\prop{A}_1}(1 + b_1 b_2) - b_1 b_2 \right)$.  Since $-1\leq\mean{\prop{A}_1}\leq1$, each term can take only three possible values $\{-3,-1,1\}$ and we again recover \eqref{eq:leggett-garg-inequality1}.  Therefore, any violation of \eqref{eq:leggett-garg-inequality1} will imply that at least one of the postulates (i-iii) of MR does not hold, or that the detector for $\prop{A}_1$ is \emph{both} invasive and ambiguous.  

We can construct many such LGIs from the same data.  For example, the three detectors in Fig.~\ref{fig:leggett-garg} allow the construction of the $2^3 - 1$ nontrivial correlation terms listed earlier, which can be combined with the three coefficients $\{-1,0,1\}$~\cite{coeff}.  Ignoring an overall sign, we can construct $(3^{2^3 - 1} - 1)/2 = 1093$ nonzero LGI correlations bounded in a similar manner to \eqref{eq:leggett-garg-inequality1}.  The subset of $(3^{2^2-1}-1)/2=13$ single-object LGIs can be obtained by ignoring the $\prop{B}_2$ detector.  Furthermore, if a fourth detector for $\prop{A}_2$ were added before the detector for $\prop{B}_2$, we could test $(3^{2^4 - 1} - 1)/2 = 7 174 453$ such LGIs.  One is formally identical to the CHSH-Bell inequality~\cite{Bell} (see also \cite{Marcovitch2010}), but tests MR and not Bell-locality.

For contrast, the original approach in \cite{Leggett} combines separate experiments for each correlation between \emph{ideal} detectors to form a single LGI.  Our approach uses a single experimental setup to determine all $2^M-1$ correlations between $M$ \emph{general} detectors to form a large number of LGIs.  Hence we obtain an exponential improvement in experimental complexity for large $M$.

\emph{Conditioned averages}.---A single-object LGI, $-3 \leq \mean{\prop{A}_1 + \prop{A}_1\prop{B}_1 - \prop{B}_1} \leq 1$, was considered in \cite{Jordan} and shown to have a one-to-one correspondence with an upper bound to the average of $\prop{A}_1$ conditioned on the positive value of $\prop{B}_1$: $\cmean{1}{\prop{A}}\leq 1$.  Three other LGIs similarly correspond to the bounds $\cmean{1}{\prop{A}}\geq -1$, and $-1 \leq \cmean{-1}{\prop{A}}\leq 1$, as checked experimentally in \cite{Goggin2010}.

We now extend these results to the two-object case using \eqref{eq:leggett-garg-inequality1}.  First we define a marginal probability of measuring $b_1$ and $b_2$ given any result of $\prop{A}_1$ as $P(b_1, b_2 | \zeta, \prop{A}_1) = \sum_{\alpha_1} P(\alpha_1 | \zeta)P(b_1, b_2 | \zeta, \alpha_1)$.  Then we define a conditional probability of measuring $\alpha_1$ given the measurement of $b_1$ and $b_2$ as, $P(\alpha_1 | \zeta, b_1, b_2) = P(\alpha_1 | \zeta)P(b_1, b_2 | \zeta, \alpha_1) / P(b_1, b_2 | \zeta, \prop{A}_1)$.  Therefore, the average of $\prop{A}_1$ conditioned on the measurements of $b_1$ and $b_2$ is $\cmean{b_1,b_2}{\prop{A}_1} = \sum_{\alpha_1} P(\alpha_1 | \zeta, b_1, b_2)\, \alpha_1$.

Using this definition, we rewrite the upper bound of \eqref{eq:leggett-garg-inequality1} as $\sum_{b_1,b_2} P(b_1, b_2 | \zeta, \prop{A}_1) \left( \cmean{b_1,b_2}{\prop{A}_1}(1 + b_1 b_2) - b_1 b_2 \right) \leq 1$ and insert the possible values for $b_1$ and $b_2$ to find the CA constraint,
\begin{equation}
  \cmean{1,1}{\prop{A}_1}\, p^+ +  \cmean{-1,-1}{\prop{A}_1}\, p^- \leq 1,
  \label{eq:conditioned-inequality1}
\end{equation}
where $p^\pm = P(\pm 1,\pm 1)/(P(1,1)+P(-1,-1))$, and $P(i,j) = P(i,j | \zeta,\prop{A}_1)$.  The degeneracy of the product value $b_1 b_2$ results in an upper bound for a \emph{convex sum} of CAs, in contrast to the single-object result in \cite{Jordan}.  A \emph{sufficient} condition for violating \eqref{eq:conditioned-inequality1} is for both CAs to exceed $1$ simultaneously.  Conversely, if all CAs were bounded by $1$, then it would be impossible to violate \eqref{eq:conditioned-inequality1} or \eqref{eq:leggett-garg-inequality1}.

\begin{figure}[tb!]
\includegraphics[scale=.5]{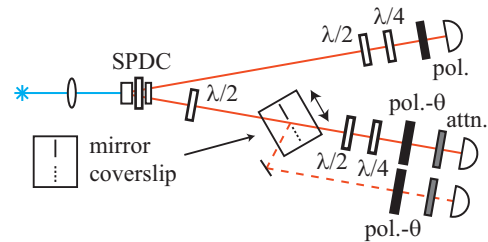}
\caption{(color online) Experimental setup. A 488 nm laser produces degenerate down-converted photon pairs. The polarization of the photon in the lower arm is rotated by $45^\circ$ with a half-wave plate, then undergoes semi-weak polarization measurement in the $\{h,v\}$ basis using Fresnel reflection ($\prop{A}_1$) that encodes the information in the resulting spatial modes, and is finally projected into the $\{\theta,\theta_\perp\}$ basis with polarizers set at angle $\theta$ ($\prop{B}_1$).  The polarization of the photon in the upper arm is projected into the $\{h,v\}$ basis with another polarizer ($\prop{B}_2$). The half and quarter waveplates prior to the polarizers are used for tomography of the input state; during data collection they are removed from the lower arm and used to switch between $h$ and $v$ polarization in the upper arm.} \label{setup}
\end{figure}

\emph{Quantum formulation}.---Projective quantum measurements produce averages of eigenvalues analogous to the results of an unambiguous detector, but non-projective quantum measurements produce averages of \emph{contextual values}~\cite{Dressel2010} which need not lie in the eigenvalue range and are therefore analogous to the results of an ambiguous detector.  By measuring $\prop{A}_1$ weakly we can find quantum mechanical violations of \eqref{eq:leggett-garg-inequality1} and \eqref{eq:conditioned-inequality1}.  

Specifically, if we start with a 2-object density operator $\op{\rho}$ and measure $\prop{A}_1$ generally such that $\op{A}_1 = \sum_{a_1} a_1 \prj_{a_1} = \sum_{\alpha_1} \alpha_1 \op{E}_{\alpha_1}$ (where $\{a_1\}$ are the eigenvalues corresponding to the projections $\{\prj_{a_1}\}$ and $\{\alpha_1\}$ are the CV corresponding to the POVM $\{\op{E}_{\alpha_1} = \op{M}_{\alpha_1}^\dagger\op{M}_{\alpha_1}\}$), and then measure $\prop{B}_1\prop{B}_2$ projectively such that $\op{B}_1\otimes\op{B}_2 = \sum_{b_1,b_2} b_1 b_2 \prj_{b_1}\otimes\prj_{b_2}$, we will find that the average correlation $\mean{C} = \mean{\prop{A}_1 + \prop{A}_1\prop{B}_1\prop{B}_2 - \prop{B}_1\prop{B}_2}$ has the form, 
\begin{equation}
  \mean{C} = \sum_{\alpha_1,b_1,b_2}P(\alpha_1; b_1,b_2 | \op{\rho})\left( \alpha_1 + \alpha_1 b_1 b_2 - b_1 b_2 \right),
  \label{eq:quantum-leggett-garg}
\end{equation}
where $P(\alpha_1; b_1,b_2 | \op{\rho}) = \Tr{\left( \op{M}^\dagger_{\alpha_1}\prj_{b_1}\op{M}_{\alpha_1}\otimes\prj_{b_2} \right)\op{\rho}}$ is the probability of measuring outcome $\alpha_1$ of the general measurement of $\prop{A}$, followed by a joint projection of $b_1 b_2$.  The appearance of the CV instead of the eigenvalues of $\op{A}$ in \eqref{eq:quantum-leggett-garg} combined with the non-separable probability $P(\alpha_1; b_1,b_2 | \op{\rho})$ allows violations of the LGI \eqref{eq:leggett-garg-inequality1}.  

The left side of \eqref{eq:conditioned-inequality1} follows from \eqref{eq:quantum-leggett-garg}, where $P(b_1,b_2|\op{\rho},\prop{A}_1) = \sum_{\alpha_1} P(\alpha_1; b_1,b_2 | \op{\rho})$ and $\cmean{b_1,b_2}{\prop{A}_1} = \sum_{\alpha_1} \alpha_1 P(\alpha_1; b_1,b_2 | \op{\rho})/P(b_1,b_2 | \op{\rho},\prop{A}_1)$ is a quantum CA as defined in \cite{Dressel2010} that converges to a \emph{weak value}~\cite{Aharonov1988} in the limit of minimal measurement disturbance.  

\begin{figure}[tb!]
\includegraphics[clip=true, viewport=.0in .0in 3.8in 1.1in, width=\columnwidth]{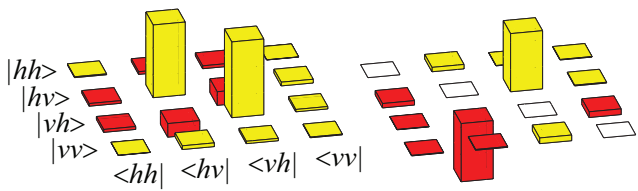}
\caption{(color online) Real (left) and imaginary (right) parts of the reconstructed density matrix in the $\{h,v\}$ basis. Yellow and red represent positive and negative values, respectively.}  \label{rho}
\end{figure}

\emph{Experimental setup}.---To implement Fig.~\ref{fig:leggett-garg} we use the polarization of an entangled biphoton with the setup shown in Fig.~\ref{setup}.  A glass microscope coverslip measures a Stokes observable $\prop{A}_1$ semi-weakly as described below, and polarizers measure Stokes observables $\prop{B}_1$ and $\prop{B}_2$ projectively.  We produce degenerate non-colinear type-II down-conversion by pumping a 2 mm walkoff-compensated BBO crystal \cite{Kwiat1995} with a narrowband 488 nm laser. The down-converted light passes through automated polarization analyzers and 3 nm bandpass filters at 976 nm in each arm before being coupled into multimode fibers connected to single photon avalanche photodiodes (SPAD). We detect coincidences using a 3 ns window. We perform state tomography with maximum likelihood estimation \cite{James2001}, which gies the state shown in Fig.~\ref{rho} with concurrence $C=0.794$, and purity $\Tr{\op{\rho}^2}=0.815$, and which resembles the pure state $\ket{\psi} = (\ket{hv} + i\ket{vh})/\sqrt{2}$.

After the state tomography, we remove the half- and quarter-wave plates from the lower arm and insert either a mirror or a coverslip using a computer-controlled translation stage. The reflected light passes though a polarization analyzer and couples into a third fiber and SPAD. We align the coverslip and the mirror to be parallel with an incidence angle of ~40$^{\circ}$ relative to the incoming beam. Finally, we optimize the fiber incoupling and balance the collection efficiencies with attenuators so that the coincidences between the upper arm and either of the lower arms differ by only a few percent when the mirror is taken in and out of the beam path.

The coverslip acts as a polarization-dependent beamsplitter measuring $\prop{A}_1 = \op{\sigma}_z$. Averaging over the 3 nm bandwidth and the thickness variation ($\sim 150\pm0.6\,\mu$m) produces an average Fresnel reflection similar to that of a single interface, with horizontal ($h$) polarization relative to the table exhibiting zero reflection near Brewster's angle and vertical ($v$) polarization exhibiting increasing reflection with incident angle. 

For a pure state of polarization $\ket{\psi}=\alpha\ket{h}+\beta\ket{v}$ with $|\alpha|^2+|\beta|^2=1$, the resulting state after passing through the coverslip is $\ket{\psi'} = (\gamma\alpha\ket{h}+\ebar\beta\ket{v})\ket{r}- (\gbar\alpha\ket{h}+\eta\beta\ket{v})\ket{t}$, where $\ket{j}$, $j\in\{t,r\}$, specify the transmitted and reflected spatial modes of the coverslip, and the reflection and transmission probabilities for $h$- and $v$-polarized light are $R_h = \gamma^2$, $R_v = \ebar^2$, $T_h = \gbar^2$, and $T_v = \eta^2$, such that $R_i+T_i=1$.  Written this way, the coverslip reflection can be viewed as a generalization of the weak measurement in \cite{Pryde2005} and discussed in \cite{Dressel2010}.  

\begin{figure}
  \includegraphics[width=\columnwidth,bb=0 0 482 155]{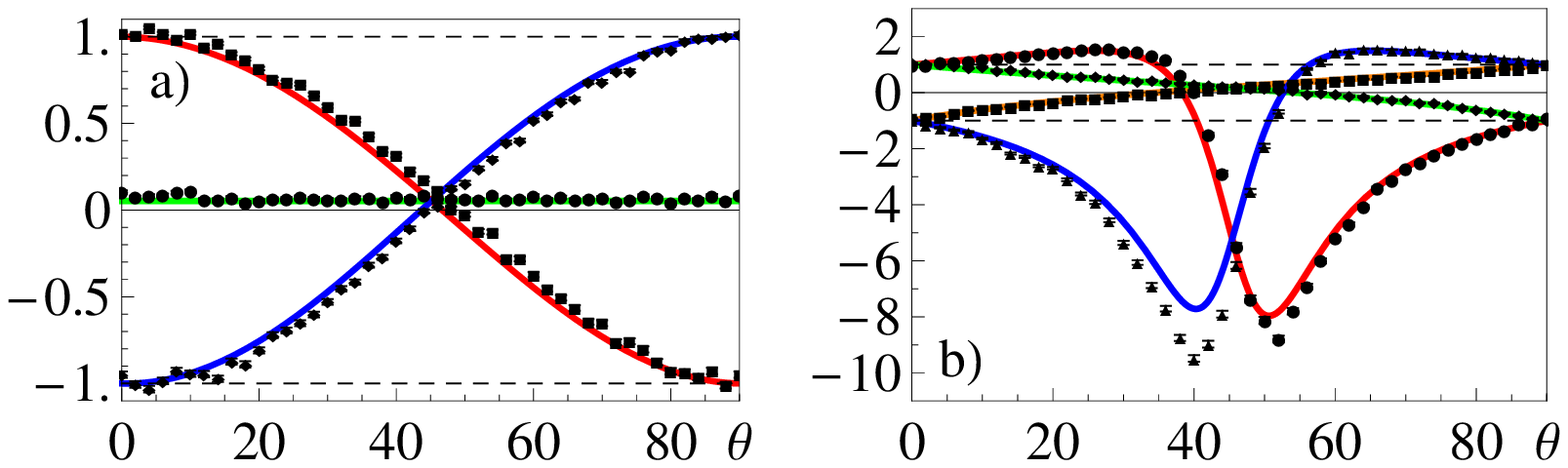} 
  \caption{(color online) In all data plots, solid lines indicate theory and points indicate experimental data.  a) $\mean{\op{\sigma}_z^{(1)}}$ (green, flat), $\cmean{\theta}{\op{\sigma}_z^{(1)}}$ (red, decreasing), and $\cmean{\theta_\perp}{\op{\sigma}_z^{(1)}}$ (blue, increasing). b) $\cmean{\theta,h}{\op{\sigma}_z^{(1)}}$ (red, bottom right), and $\cmean{\theta_\perp,v}{\op{\sigma}_z^{(1)}}$ (blue, bottom left), violating negative bounds, unlike $\cmean{\theta_\perp,h}{\op{\sigma}_z^{(1)}}$ (orange, increasing), and $\cmean{\theta,v}{\op{\sigma}_z^{(1)}}$ (green, decreasing).}  
  \label{condav}
\end{figure}

From $\ket{\psi'}$, we find the measurement operators for the back-action of the coverslip outcomes to be $\op{M}_r = \gamma\prj_h+\ebar\prj_v$ and $\op{M}_t = \gbar\prj_h+\eta\prj_v$, where $\prj_i$, $i\in\{h,v\}$, are polarization projectors.  The corresponding POVM elements are $\op{E}_r = R_h\prj_h+R_v\prj_v$ and $\op{E}_t = T_h\prj_h+T_v\prj_v$, with which we can expand the polarization Stokes operator as $\op{\sigma}_z = \prj_h-\prj_v = \alpha_r \hat{E}_r + \alpha_t \hat{E}_t$, as discussed before \eqref{eq:quantum-leggett-garg}, where $\alpha_r=(T_h+T_v)/(R_h-R_v)$ and $\alpha_t=-(R_h+R_v)/(R_h-R_v)$ are the CV.

We determine the values of $R_h$ and $R_v$ with calibration polarizers before the coverslip, yielding $R_h=0.0390\pm0.0007$ and $R_v=0.175\pm0.001$.  The reflected arm is largely projected to $v$, while the transmitted arm is only weakly perturbed, making the total coverslip effect a \emph{semi-weak measurement}.  The CV, $\alpha_r = -13.1\pm0.1$ and $\alpha_t = 1.57\pm0.01$, are correspondingly amplified from the eigenvalues of $\op{\sigma}_z$.

\emph{Results}.---To complete the state preparation, we place a half-wave plate before the coverslip in the lower arm and rotate the polarization by $45^\circ$ to produce a state similar to $\ket{\psi''}=(\ket{ha}+i\ket{vd})/\sqrt{2}$. We then measure \eqref{eq:leggett-garg-inequality1} by choosing the observables $\prop{A}_1$, $\prop{B}_1$, and $\prop{B}_2$ to be the Stokes observables $\op{\sigma}_z^{(1)}$, $\op{\sigma}_\theta^{(1)}$ and $\op{\sigma}_z^{(2)}$, respectively, where $\op{\sigma}_\theta$ is the $\op{\sigma}_z$ operator rotated to the $\{\theta,\theta_\perp\}$ basis (e.g.  $\op{\sigma}_{0^\circ} = \op{\sigma}_z$ and $\op{\sigma}_{45^\circ} = \op{\sigma}_x$).  By changing the single parameter, $\theta$, we can explore a range of observables.

\begin{figure}
  \includegraphics[scale=.51,bb=0 0 300 198]{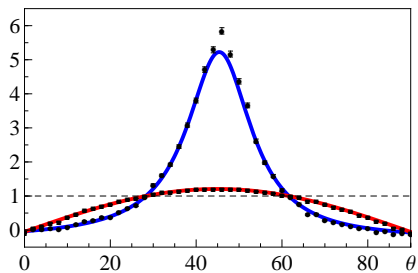}
  \caption{(color online) LGI correlation $\mean{-\sigma_z^{(1)} - \sigma_z^{(1)}\sigma_\theta^{(1)}\sigma_z^{(2)} - \sigma_\theta^{(1)}\sigma_z^{(2)}}$ (red, squares) and the corresponding convex sum of the CAs $\cmean{\theta,h}{-\op{\sigma}_z^{(1)}}$ and $\cmean{\theta_\perp,v}{-\op{\sigma}_z^{(1)}}$ (blue, circles), both violating their upper bounds of 1 in the same domain of $\theta$.  Compare to Fig.~\ref{condav} (b) and note that the LGI violation includes the region where the two CAs both exceed their bounds.}
  \label{calg}
\end{figure}

Fig.~\ref{condav} shows the various averages of $\op{\sigma}_z^{(1)}$.  Averaging all results for orthogonal settings on $\op{\sigma}_\theta^{(1)}$ and $\op{\sigma}_z^{(2)}$ gives the expectation value $\mean{\op{\sigma}_z^{(1)}}$, which is properly constant and near zero for all $\theta$ since the reduced density operator is almost fully mixed.  Averaging only the results for the orthogonal settings of $\op{\sigma}_z^{(2)}$ gives the single CAs $\cmean{\theta}{\op{\sigma}_z^{(1)}}$ and $\cmean{\theta_\perp}{\op{\sigma}_z^{(1)}}$, which are also well-behaved.  Finally, averaging only the results for specific settings gives the double CAs $\cmean{\theta,v}{\op{\sigma}_z^{(1)}}$, $\cmean{\theta_{\perp},h}{\op{\sigma}_z^{(1)}}$, $\cmean{\theta,v}{\op{\sigma}_z^{(1)}}$, and $\cmean{\theta_{\perp},v}{\op{\sigma}_z^{(1)}}$, which can exceed the eigenvalue range for some range of $\theta$ due to the non-local correlations in the entangled biphoton state.

Using the same set of data, Fig.~\ref{calg} shows the upper bound of the LGI $-3 \leq \mean{-\sigma_z^{(1)} - \sigma_z^{(1)}\sigma_\theta^{(1)}\sigma_z^{(2)} - \sigma_\theta^{(1)}\sigma_z^{(2)}} \leq 1$ being violated in the same range of $\theta$ that the appropriate convex sum of $\cmean{\theta,h}{-\op{\sigma}_z^{(1)}}$ and $\cmean{\theta_\perp,v}{-\op{\sigma}_z^{(1)}}$ violates its upper bound according to \eqref{eq:conditioned-inequality1}.

\begin{figure}[t!]
  \includegraphics[scale=.53,bb=0 0 300 193]{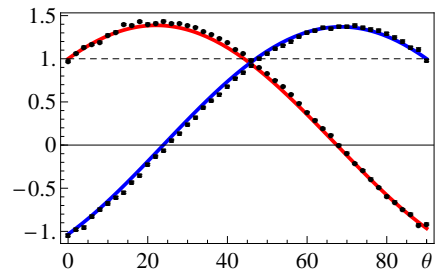}
  \caption{(color online) LGI correlations $\mean{\sigma_z^{(1)}\sigma_z^{(2)} + \sigma_z^{(2)}\sigma_\theta^{(1)} - \sigma_z^{(1)}\sigma_\theta^{(1)}}$ (red, circles), and $\mean{-\sigma_z^{(1)}\sigma_z^{(2)} + \sigma_z^{(2)}\sigma_\theta^{(1)} + \sigma_z^{(1)}\sigma_\theta^{(1)}}$ (blue, squares) violating their upper bounds of 1 for nearly the entire $\theta$ domain.} 
  \label{twolg}
\end{figure}

We can violate several more LGIs using the same set of data as well.  Fig.~\ref{twolg} shows two such correlations, $\mean{\sigma_z^{(1)}\sigma_z^{(2)} + \sigma_z^{(2)}\sigma_\theta^{(1)} - \sigma_z^{(1)}\sigma_\theta^{(1)}}$, and $\mean{-\sigma_z^{(1)}\sigma_z^{(2)} + \sigma_z^{(2)}\sigma_\theta^{(1)} + \sigma_z^{(1)}\sigma_\theta^{(1)}}$ that between them violate an upper bound over nearly the whole range of $\theta$, for illustration.

All solid curves in Figures~\ref{condav},~\ref{calg}, and~\ref{twolg} are quantum predictions analogous to \eqref{eq:quantum-leggett-garg} using the measurement operators, CV, and the reconstructed initial state.  They also include compensation for a few percent deviation in the thickness of the half-wave plate in the upper arm.  The points indicate experimental data and include Poissonian error bars.  The small discrepancies between theory and data can be explained by sensitivity to the state reconstruction and additional equipment imperfections.  The violations indicate either that MR is inconsistent with experiment or that the semi-weak measurement device is \emph{both} invasive and ambiguous in the MR sense.

\emph{Conclusion}.---We have illustrated the derivation of generalized single-setup LGIs allowing for multiple particles and measurements with more realistic detectors by considering a two-particle example, and have demonstrated simultaneous violations of several such two-party LGIs using the same data set from a biphoton polarization experiment.  Due to the single setup, any dataset may be similarly examined for inherent LGI violations.

\begin{acknowledgments}
This work was supported by the NSF Grant No. DMR-0844899, ARO Grant No. W911NF-09-1-0417, and a DARPA DSO Slow Light grant.
\end{acknowledgments}


\end{document}